\documentclass[useAMS,usenatbib,usegraphicx]{mn2e}

\title[H$\alpha$ long term monitoring of the Be star $\beta$~Cep~Aa]
  {H$\alpha$ long term monitoring of the Be star $\beta$~Cep~Aa}
\author[G. Catanzaro]
  {G.~Catanzaro\thanks{E-mail: Giovanni.Catanzaro@oact.inaf.it}\\
  INAF - Osservatorio Astrofisico di Catania, Via S. Sofia 78, 95123 Catania, Italy}
 \date{Accepted 2008 March 27.  Received 2008 March 27; in original form 2008 February 12}

\pagerange{\pageref{firstpage}--\pageref{lastpage}} \pubyear{2008}

\def\LaTeX{L\kern-.36em\raise.3ex\hbox{a}\kern-.15em
    T\kern-.1667em\lower.7ex\hbox{E}\kern-.125emX}

\begin{document}

\label{firstpage}

\maketitle

\begin{abstract}
Recent papers published in the last years contributed to resolve the enigma 
on the hypothetical Be nature of the hot pulsating star $\beta$~Cep. This star 
shows variable emission in the H$\alpha$ line, typical for Be stars, but its 
projected rotational velocity is very much lower than the critical limit, 
contrary to what is expected for a typical Be star. The emission has been 
attributed to the secondary component of the $\beta$~Cep spectroscopic binary 
system.

In this paper, using both ours and archived spectra, we attempted to recover
the H$\alpha$ profile of the secondary component and to analyze its behavior 
with time for a long period. To accomplish this task, we first derived the 
atmospheric parameters of the primary: T$_{\rm eff}$\,=\,24000\,$\pm$\,250 K 
and $\log g$\,=\,3.91\,$\pm$\,0.10, then we used these values to compute its 
synthetic H$\alpha$ profile and finally we reconstructed the secondary's 
profile disentangling the observed one.

The secondary's H$\alpha$ profile shows the typical two peaks emission
of a Be star with a strong variability. We analyzed also the behavior
versus time of some line width parameters: equivalent width, V/R, FWHM, peaks
separation and radial velocity of the central depression.

Projected rotational velocity ($v \sin i$) of the secondary and the dimension 
of the equatorial surrounding disk have been estimated, too.

\end{abstract}


\section{Introduction}
The star $\beta$ Cephei (HD\,205021, V\,=\,3.2) is well known to be the 
prototype of a class of hot pulsating stars. Several studies reported in 
the recent
literature have been outlined a star oscillating with at least five 
frequencies: $f_1$\,=\,5.2497104~c~d$^{-1}$, 
$f_2$\,=\,5.385~c~d$^{-1}$ and $f_3$\,=\,4.920~c~d$^{-1}$ \citep{aerts94}
and $f_4$\,=\,5.083~c~d$^{-1}$ and $f_5$\,=\,5.417~c~d$^{-1}$
\citep{telting97}.

Further, the star ($\beta$~Cep~A) is actually a complicated multiple system.
In fact, beside to have a visual companion ($\beta$~Cep~B, V\,=\,7.9) at a 
distance of 13.4'', it is also a member of a spectroscopic system whose 
second star ($\beta$~Cep~Aa) was discovered using speckle interferometry
by \citet{gezari72} at a distance of  $\approx$0.25''. The parameters of the close 
binary orbit have been determined later by \citet{pigu92} combining speckle 
interferometry and the variations in the pulsation period, due to the 
so-called light time effect. Recent speckle measurements by \citet{hartkopf01} 
place the position of the companion to about 0.1'' from the primary and \citet{balega02}
found a magnitude difference of 1.8 between the two components in a red filter 
centered on 810~nm and 60~nm wide. More observations are still necessary to 
improve the orbital solution.

Another characteristic that gives more interest to this object is the variable 
emission observed in the H$\alpha$. Since \citet{karpov33}, the presence
of this emission had been reporting by several authors: 
\citet{wilson56} or \citet{kaper95}, who observed a decreasing
emission from 1990 to 1995. In the same years, $\beta$~Cep has been observed
with the coud\'e spectrograph of the 2.6 meter telescope of the Crimean 
Astrophysical Observatory by \citet{panko97}. Over the entire period covered 
by their observations, the H$\alpha$ was in emission with a pronounced
two component structure and a small intensity above the continuum level.
The same authors stated that in November 1987 the line showed an absorption
profile without any sign of emission.

The kind of variable emission discussed so far is typically the characteristic 
of Be stars. Be stars are rapidly rotating B-type stars that lose mass in an 
equatorial, circumstellar disk (see the recent review by \citet{porter03}).
A general characteristic of Be stars is that they rotate very fast, typically
at about 70$\%$-80$\%$ of their critical limit (v$_{\rm crit}$\,=\,$\sqrt{GM_*/R_*}$) 
corresponding to a several hundred of km s$^{-1}$ \citep{slett82}. On the contrary,
the projected rotational velocity of $\beta$~Cep~A (26~km~s$^{-1}$, \citet{morel06})
is very much lower than this limit and this questions its real nature. 

\citet{hadrava96} separated emission and absorption components of the H$\alpha$
profile from their 1996 spectra and found that they move in antiphase. They 
speculated on two possible scenarios: {\it i)} the observed emission arises 
from reprocessing of the light originating in the stellar photosphere in an 
outer envelope, {\it ii)} the star is not a pulsating object but a rapidly rotating 
oblique rotator with a magnetic dipole geometry. 

Recently, \citet{schnerr06} argued for the first time that the H$\alpha$
emission observed from the $\beta$~Cep system is not related to the slowly rotating
primary star, but to the secondary, being the latter a classical Be star. 
Unfortunately, they obtained only one spectrum at the {\it Nordic Optical
Telescope} and, then, no conclusion about the variability of the emission could 
be drawn.

In this paper we try to recover the recent emission history of $\beta$~Cep~Aa.
To achieve this goal, we firstly disentangled the spectra of the
two components from the observed one. This has been done using a 
synthetic profile for the H$\alpha$ of $\beta$~Cep~A, for which T$_{\rm eff}$ 
and $\log g$ have been estimated as described in Sect.~\ref{primary}.
Then, we analyzed the behavior with time of some H$\alpha$ width parameters like 
its equivalent width, the full width at half maximum, the peaks separation and 
the radial velocity of the central depression. Finally, we attempted to estimate
the $v \sin i$ of $\beta$~Cep~Aa and the dimension of its surrounding disk.

\section{Observation and data reduction}
\label{observ}
We observed $\beta$~Cep for more than one year from July 2005 to November 2006. 
To extend our coverage to the past years, we queried various astronomical archives 
on the internet. At the end, we collected
a number of 72 spectra covering the period from September 1993 to November 2006,
for a total of more than 13 years. In the following we detailed each single
observatory and instrument:

\begin{itemize}

\item 42 spectra have been taken from the archive of the 
     {\it Ritter Observatory}, University of Toledo, OH USA, equipped 
      with 1-m Ritchey-Chretien reflector. 
      The echelle spectrograph is connected to the 
      Cassegrain focus of the telescope by fiber optic cable of 200~$\mu$m diameter. 
      The detector is a CCD manufactured by EEV with 1200 x
      800 pixels of 22.5~$\mu$m size. The spectra used in this paper have been
      observed with R\,$\approx$\,26000; 

\item 4 spectra have been downloaded from the {\it Isaac Newton Telescope}
      archive, three of them have been acquired with IDS spectrograph and
      one with MUSICOS. The lines of the Th-Ar lamp show that for all the
      spectra the resolution is R\,$\approx$\,33000;

\item 7 spectra have been downloaded from the archive of the 
      Elodie@OHP~1.93~m telescope. Spectra have been reduced by the standard 
      pipeline procedure, described in \citet{baranne96}. Resolving power of
      these spectra is $\approx$\,42000;  

\item the 91 cm telescope of the {\it INAF\,-\,Osservatorio 
      Astrofisico di Catania}, has been used by us to carried out 19 spectra
      of our target. The telescope is fiber linked to a REOSC echelle
      spectrograph, which allows to obtain R\,=\,20000 spectra in the range 
      4300-6800 {\AA}. The resolving power has been checked using emission 
      lines of the Th-Ar calibration lamp. Spectra were recorded on a thinned, 
      back-illuminated (SITE) CCD with 1024 x 1024 pixels of 24~$\mu$m size, 
      typical readout noise of 6.5 e$^{-}$ and gain of 2.5 e$^-$/ADU. These spectra
      have been extracted from a more complete set of observations \citep{catanzaro08},
      in such a way to choose the one with the best SNR for each observing night.

The stellar spectra, calibrated in wavelength and with the continuum normalized
to a unity level, were obtained using standard data reduction procedures for 
spectroscopic observations within the NOAO/IRAF package, that is: bias frame 
subtraction, trimming, scattered light correction, flat-fielding, fitting 
traces and orders extraction and, finally, wavelength calibration. IRAF 
package {\it rvcorrect} has been used to include the velocity correction 
due to the Earth's motions, all the spectra were then reduced into the 
heliocentric rest of frame.

\section{Atmospheric parameters of the primary}
\label{primary}
The first task we had to asses in our study was to compute the synthetic 
H$\alpha$ profile of the primary component that we used in the reconstruction
process of the secondary's  profile. 
This calculation is possible if one knows the effective
temperature and surface gravity of the primary component of the system.

The approach we used in this study to determine T$_{\rm eff}$ and $\log g$
was to compare the observed and theoretical profiles of H$\gamma$ and H$\beta$
lines by minimizing the goodness-of-fit parameter:\\

$\chi^2 = \frac{1}{N} \sum \left(\frac{I_{\rm obs} - I_{\rm th}}{\delta I_{\rm obs}}\right)^2$

\bigskip

\noindent
where $N$ is the total number of points, $I_{\rm obs}$ and $I_{\rm th}$ 
are, respectively, the intensities -- normalized to the continuum -- of the observed
and computed profiles and $\delta I_{\rm obs}$ is the photon noise. Errors in
T$_{\rm eff}$ and $\log g$ are estimated
as the variation in the parameters which increases the $\chi^2$ of a unit.

\end{itemize}
\begin{figure}
\includegraphics[width=9cm]{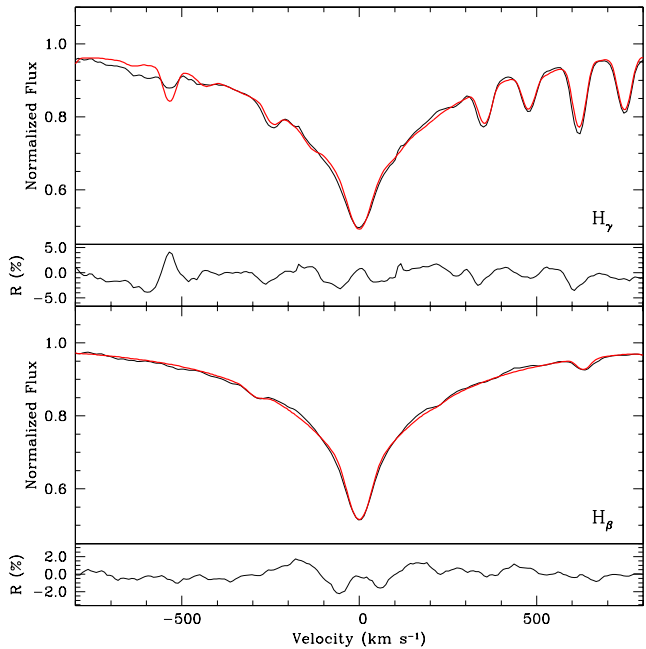}
\caption[]{Comparison between observed (black) and synthetic (red) profile of 
H$\gamma$ and H$\beta$ lines. Atmospheric parameters used for this calculation are:
T$_{\rm eff}$\,=\,24000\,$\pm$\,250 K and $\log g$\,=\,3.91\,$\pm$\,0.10.
For each Balmer line we reported also the residuals (O\,-\,C) expressed in percentage.
Note the strong oxygen lines present in the wings of the H$\gamma$ line.}
\label{balmer}
\end{figure}

The observed profiles presented in Fig.~\ref{balmer} have been calculated by 
averaging a number of 25 spectra collected by us during the Julian Day 
2\,554\,041.45 \citep{catanzaro08}. Doppler correction has been 
applied to remove any radial velocity shift due to pulsations. This average
spectrum has been used in order to improve the signal-to-noise ratio of
each single exposures, in particular we obtained SNR\,$\approx$\,250 in the
continuum close to the H$\gamma$ and SNR\,$\approx$\,300 in the continuum
close to the H$\beta$.

Theoretical profiles have been computed with SYNTHE\footnote{All 
the Kurucz codes (ATLAS9 and SYNTHE) have been used in the Linux 
version implemented by \citet{sbord04}}
\citep{kur81} on the basis of ATLAS9 \citep{kur93}
atmosphere models.  To reduce the number of free parameters, the rotational
velocity of $\beta$~Cep~A has been determined fitting the line profiles of the 
Si{\sc iii} triplet at $\lambda$\,4552, 4567 and 4574 {\AA}.
The value obtained of $\rm v \sin i$\,=\,30~$\pm$~1~km~s$^{-1}$ is
almost coincident with the line broadening of 29~km~s$^{-1}$ measured by 
\citet{morel06} in their analysis of high resolution spectra (R\,=\,50\,000).

\begin{figure}
\includegraphics[width=9cm]{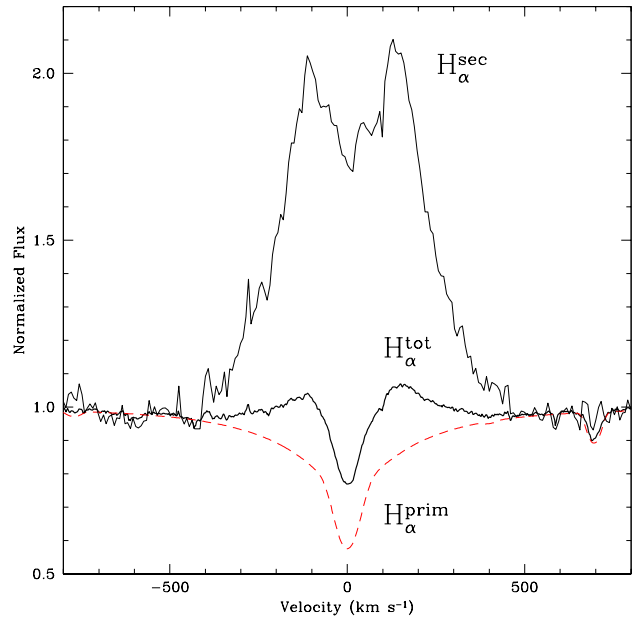}
\caption[]{In this figure we illustrated the method used to recover the 
H$\alpha$ profile of the secondary. With bold solid line we represented
the observed composite spectrum, with red dashed line we represented
the synthetic primary component and with the normal line we indicated
the reconstructed secondary profile.}
\label{example}
\end{figure}

The best fit was achieved for the following adopted parameters: 
T$_{\rm eff}$\,=\,24000\,$\pm$\,250 K and $\log g$\,=\,3.91\,$\pm$\,0.10.
In Fig.~\ref{balmer} we showed the comparison between the observed and 
computed lines profiles together with their residuals (O~-~C). It is interesting
to note the two symmetrical bumps that appear both on the blue and red side
of H$\beta$, approximately at $\pm$\,150~km~s$^{-1}$ from the core of the line.
They are located at the same velocity of the peaks observed in the H$\alpha$
of $\beta$~Cep~Aa (see Fig.~\ref{example}), then they could be connected to the H$\beta$
of the secondary component. On the other side, $\beta$~Cep~A is a non-radial
pulsator, thus the peaks could be the result of the average process we undertook
to recover the mean profiles used for T$_{\rm eff}$ and $\log g$ determination.
This distortion, whatever its origin may be, could led to a maximum error on the parameters
of the order of 2~$\%$.

\begin{figure*}
\includegraphics[width=18cm]{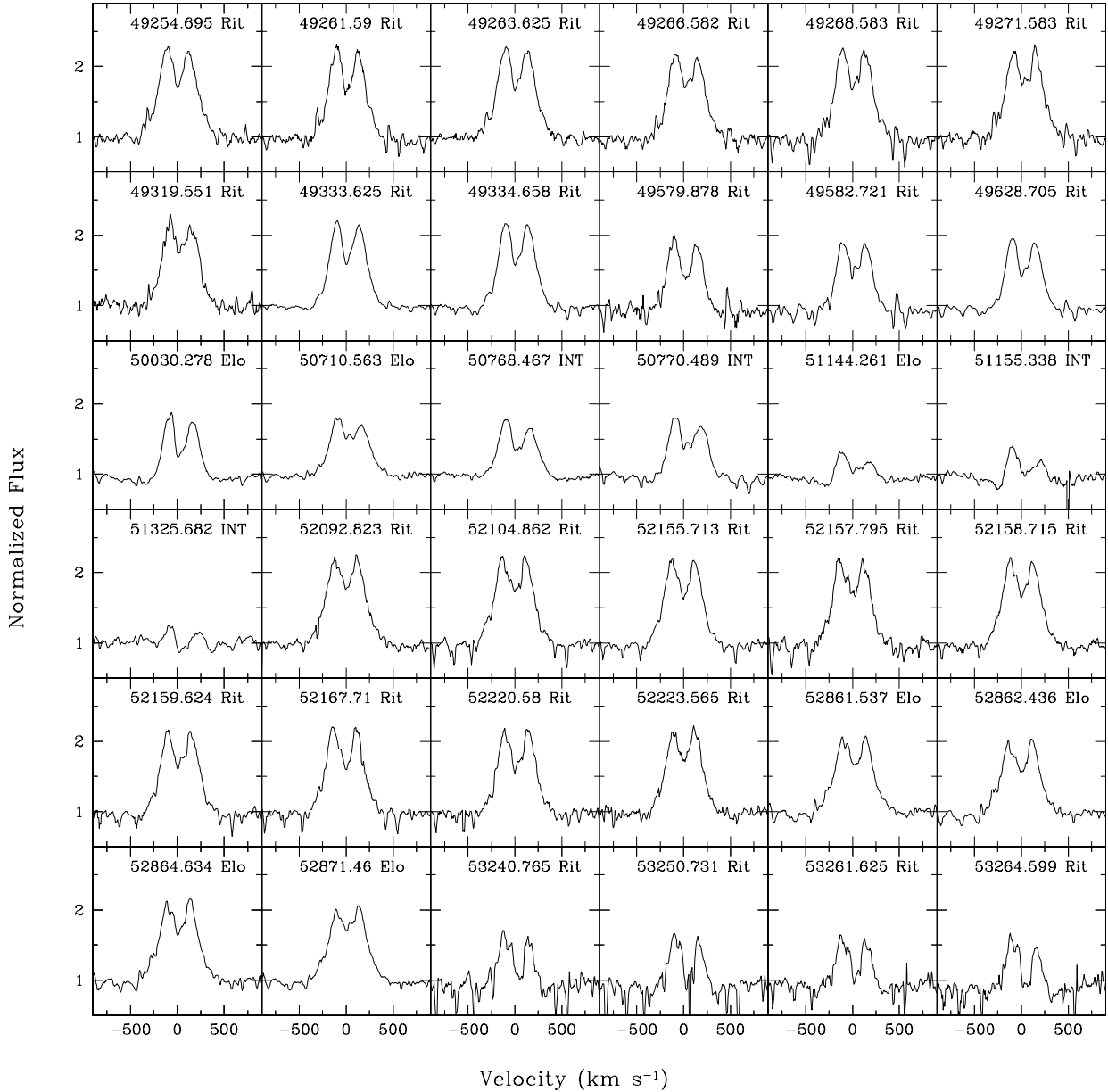}
\caption[]{H$\alpha$ profiles of $\beta$~Cep~Aa reconstructed with the procedure 
described in the text. In each box we reported the heliocentric julian day of the
observation and a label indicating the observatory: Rit - Ritter Observatory,
INT - Isaac Newton Telescope, Elo - Elodie and OAC - Catania Astrophysical 
Observatory.}
\label{profiles1}
\end{figure*}

\begin{figure*}
\includegraphics[width=18cm]{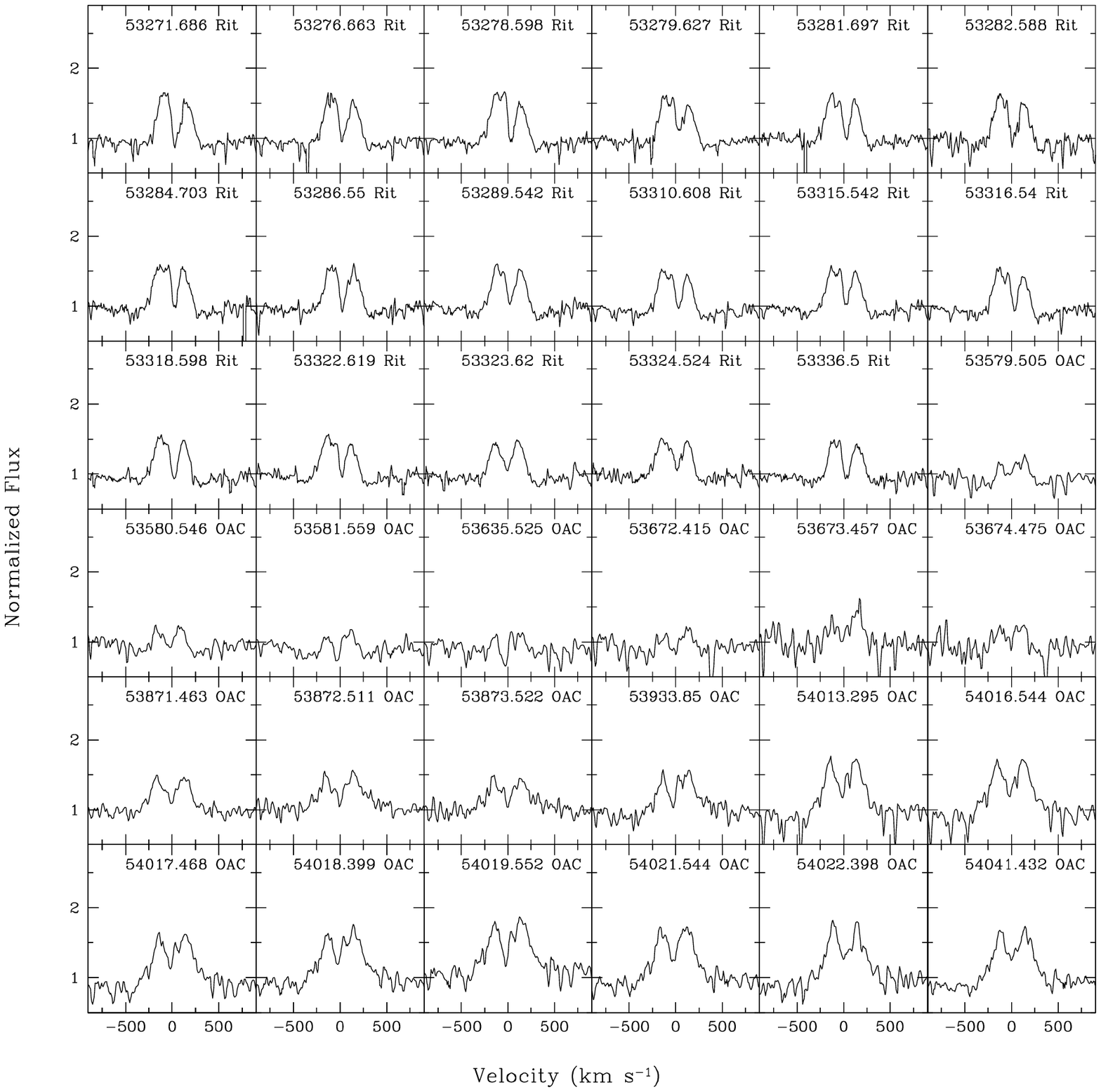}
\caption[]{As in Fig.~\ref{profiles1}}
\label{profiles2}
\end{figure*}

In the particular case of our target, this procedure could be considered valid
under the initial hypothesis that H$\gamma$ and H$\beta$ are not influenced
by the secondary component of the binary system. As a check on the consistency 
of our T$_{\rm eff}$ and $\log g$, we estimated the abundance of silicon from
lines of different ionization states. In particular we used the Si{\sc iii}
triplet and the Si{\sc iv} $\lambda$4654 {\AA}. The abundance has been
computed by means of spectral synthesis analysis using the microturbulence
found by \citet{gies92}, that is: 7.6\,$\pm$\,1.5~km~s$^{-1}$. We derived 
$-$4.67\,$\pm$\,0.08 from Si{\sc iii} and $-$4.65\,$\pm$\,0.10 from Si{\sc iv}. 
The consistency between these two determinations
convinced ourselves about the accuracy of our atmospheric parameters.
As adopted Si abundance we computed the weighted average between the two previous 
values, that is $\log$\,N(Si)/N$_{\rm tot}$\,=$-$4.66\,$\pm$\,0.06. This 
abundance, that could be expressed as $\log \epsilon(Si)$\,=\,7.38$\pm$\,0.06\footnote{This 
conversion has been calculated considering normal helium abundance.},
is fully consistent with typical abundances for B-type field stars reported
in various literature study: for instance, \citet{dufton90} gave 
$\log \epsilon(Si)$\,=\,7.5\,$\pm$\,0.2 for both the OB associations
$\it h$ and $\chi$ Per and Cep~OB3, \citet{gies92} found 
$\log \epsilon(Si)$\,=\,7.58 from their sample of B stars and \citet{cunha94} found 
7.40\,$\pm$\,0.15 in a sample of B stars in the Orion association.  

For the sake of comparison, we searched for other determinations of T$_{\rm eff}$ 
and $\log g$ in the literature. In particular, \citet{heyn94} found 
T$_{\rm eff}$\,=\,24550 and $\log g$\,=\,3.772 and \citet{niec05} derived 
T$_{\rm eff}$\,=\,24150\,$\pm$\,350 K and $\log g$\,=\,3.69. As the reader can
see, despite effective temperatures are perfectly consistent with our value, 
our gravity is $\approx$\,0.2 dex greater than previous determinations.
Anyway, since other authors did not report their errors on $\log g$, no conclusion
could be drawn about the consistency between our and literature values.

\section{Extraction of the secondary spectrum}

The procedure we undertook to recover the H$\alpha$ profiles of the 
secondary component could be summarized as follows:

\begin{itemize}

\item first, we computed synthetic H$\alpha$ profiles for the primary
component using the atmospheric parameters derived in the previous section.
Even for this line we used ATLAS9 and SYNTHE;

\item then, we recovered the secondary H$\alpha$ line profiles from the
following formula:

\begin{equation}
F_{\rm tot}\,=\,\frac{F_{\rm prim}~r~+~F_{\rm sec}}{1~+~r}
\label{ftot}
\end{equation}

\noindent
where $F_{\rm tot}$ is the observed profile, $F_{\rm prim}$ is
the synthetic H$\alpha$ profile of the primary, $r$ is the luminosity ratio
between the two components computed from the magnitude difference
$\Delta\,M$\,=\,1.8 \citep{balega02} opportunely scaled at the H$\alpha$ wavelength.
Then from Eq.~\ref{ftot} it follows:

\begin{equation}
F_{\rm sec}\,=\,(1~+~r)~F_{\rm tot}~-~F_{\rm prim}\,r
\end{equation}

\end{itemize}

Since the primary component is a pulsating star with an amplitude of
about 30 km~s$^{-1}$ (\citet{aerts94}, \citet{telting97}, \citet{catanzaro08}), 
it has been necessary to correct the synthetic spectrum for the radial velocity 
corresponding to the pulsating phase. For this aim we used the 
nearby C{\sc ii} doublet at $\lambda \lambda$ 6578, 6582 {\AA}, both belonging
to $\beta$~Cep~A. For the sake of clarity, we show in Fig.~\ref{example} 
an example of our procedure.

The profiles of $\beta$~Cep~Aa, recovered as described before and
converted in the heliocentric velocity scale, are showed in 
Fig.~\ref{profiles1} and in Fig.~\ref{profiles2}. Inspection of those
two figures revealed the strong variability of $\beta$~Cep~Aa, even though
no transition between shell into normal B spectrum has been observed. It
always shows a double-peak symmetrical profile, with V/R $\approx$~1.
Spectra are ordered by Julian Date. 

\begin{figure}
\includegraphics[width=13cm,height=17cm]{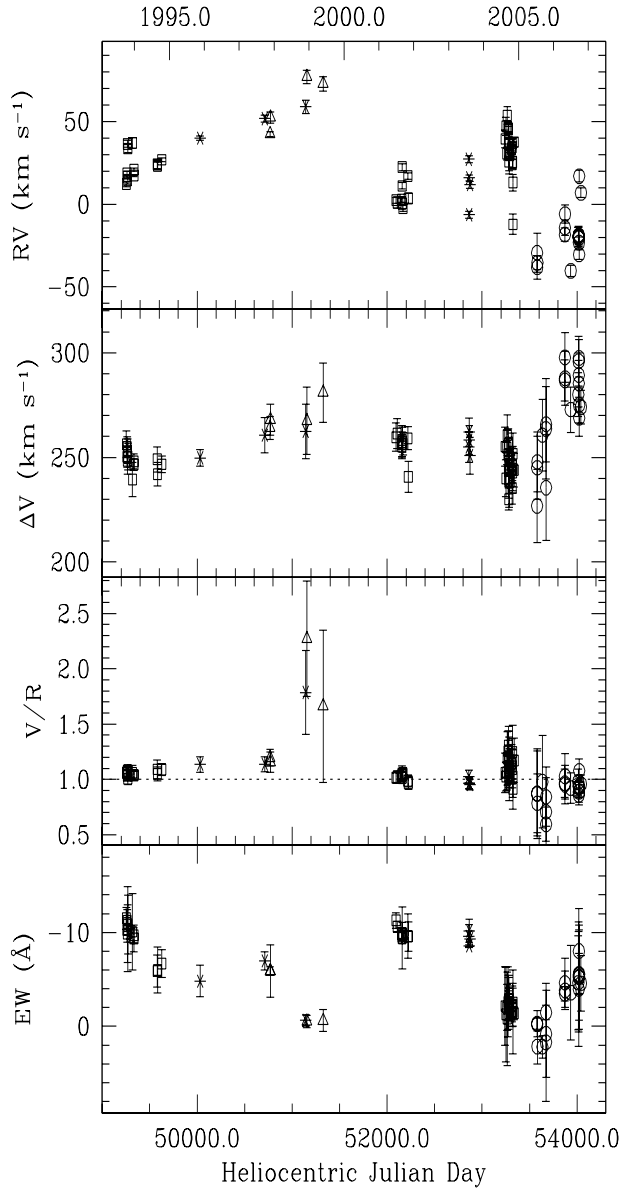}
\caption[]{Variability of some quantities characteristics of the H$\alpha$ 
emission profile plotted in function of the Heliocentric Julian Day. From bottom to
top panel: equivalent width expressed in {\AA}, V/R, peaks separation
$\Delta$V and radial velocities of the central depression, all 
expressed in km~s$^{-1}$. Meaning of the symbols
is: {\it squares} - data from Ritter Observatory, 
{\it triangles} - data from INT, {\it asterisks} - data from Elodie 
and {\it circles} - data from Catania Astrophysical 
Observatory. In the EW panel, the H$\alpha$ emission increase going up; the 
dotted line in the V/R panel represents the V\,=\,R symmetry. The top axis
is labeled in fraction of years.}
\label{variab}
\end{figure}

\section{H$\alpha$ Variability and disk properties}
In this section we analyzed the behavior with time of some line width 
parameters of the H$\alpha$ emission profile, i.e.: equivalent width (EW), 
the full width at half-maximum (FWHM), the peaks separation ($\Delta V$), 
the radial velocity of the central depression (RV$_{\rm CD}$) and the ratio 
of blue to red peak intensities (V/R) computed after subtraction of the
continuum level\footnote{Since our spectra have been normalized to the
unity, we computed this ratio as (V-1)/(R-1).}. In order to better measure 
positions and flux levels of blue and red 
peaks, we fitted each profile with a function defined as the sum of the two
gaussians centered respectively on the blue and the red peak. A similar fit 
has been considered to compute the radial velocity of the central depression.
All this data with their relative errors have been reported in 
Tab.~\ref{journal}.

In Fig.~\ref{variab} we plotted all these quantities versus time.
As a general consideration, we observed two different regimes of 
variations: a smoothed one between Julian Date 2449250 and 2453000 
and a more rapid variation starting from JD\,$>$\,2453000.

\begin{figure}
\includegraphics[width=9cm]{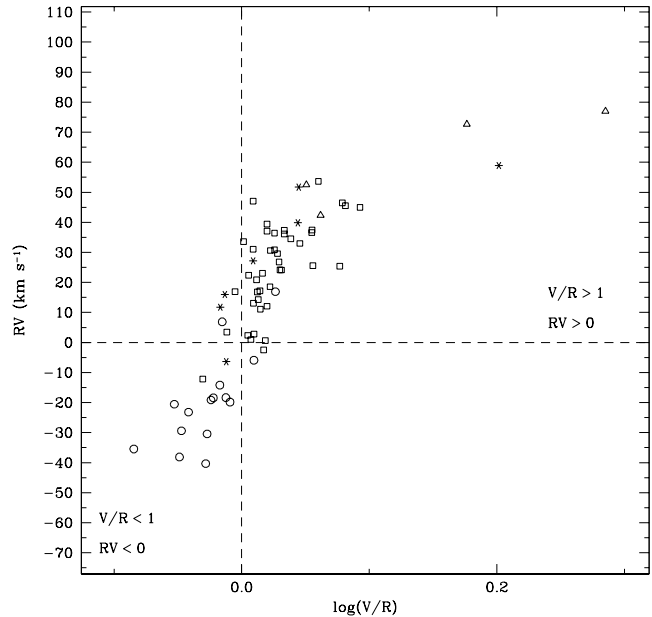}
\caption[]{Correlation between V/R, expressed in logarithm,  and radial velocity 
of the central depression. }
\label{vr_rv}
\end{figure}

In the bottom panel of Fig.~\ref{variab} we reported the equivalent width 
of H$\alpha$ versus time. By definition, it is straightforward that lines
with emission profiles will have negative EW. In the plot we inverted
the directions of y axis in such a way to have ascending curve in
correspondence of greater emission. EW decreases from September 1993 to 
May 1999, then a new strong emission phenomenon occurs in July 2002 that 
remains constant till August 2003. Again EW decreases, it reaches a minimum 
during September 2005 and it increases again until November 2006, date of last 
spectrum of our series.

As concerns the FWHM we found its value almost constant in time, 
about 400~$\pm$~30~km~s$^{-1}$, at least at the resolving power of our spectra. 

The peaks separation reveled an increase from September 1993 to May 1999, then
it decreased reaching a minimum in June 2005. After that date, the separation
begins to increase again (third panel from bottom of Fig.~\ref{variab}). 

The fourth panel shows the behavior of the radial velocity of the central 
depression. It shows a progressive red shift till May 1999. Then, we observed
a blue shift till June 2005. After this date velocities present much more
scatter, probably due to a worse quality of data. In general, from the 
variations of the radial velocity, one could
draw some useful indication about the state of the disk: expanding, contracting
or stationary. This is true provided that the radial velocity of the underlying
star is well known. Unfortunately, since this is not possible with our data,
we could not say anything about the equilibrium state of the envelope.

A similar trend has been observed for the V/R ratio, but with an exceptional 
increase (V/R~$>$~1.5) for three julian dates: 51144.261 (Elodie),
51155.338 (INT) and 51325.682 (INT). Cyclic variability of V/R has already
been observed in a number of Be stars \citep{hanu95} and this evidence has 
been theoretically explained in terms of one-armed density
waves precessing around the central star, the so-called Okazaki's model
\citep{okazaki91}. When the dense part is on the
approaching side of the disk, the V peak is higher than R. On the contrary,
when dense part is in the receding part of the disk, R peak will be
enhanced. 

It is well known from literature (see for example \citet{hubert87}), that in 
addition to the V/R variation Be stars are always accompained by profile shifts:
blueward when red component is stronger (V/R\,$<$\,1), redward when blue
component is stronger (V/R\,$>$\,1). This feature is again explained in term of
the one-armed density model as proposed by \citet{okazaki96}. In Fig.~\ref{vr_rv}
we show the correlation between RV and V/R. It is clear that when V/R\,$>$\,1
the profile is redshifted (RV\,$>$\,0) and, on the contrary, the profile
is blueshifted (RV\,$<$\,0) when V/R\,$<$\,1.

For all these quantities, we attempted to search for periodic variability
using the {\it Phase Dispersion Method (PDM)} \citep{stellin78} 
as coded in the NOAO/IRAF package, but no succesfully results have been achieved.

\citet{dachs86} and \citet{hanu86}, studying the kinematics of the disks 
surrounding Be stars, found some useful correlations between line width 
parameters (in particular, FWHM and peak separation) and stellar
projected rotational velocity, $v \sin i$. Later, the existence of those 
correlations have been confirmed by \citet{hanu88}. Using those correlations,
we attempted to derive the $v \sin i$ of the underlying star. In particular,
\citet{hanu88} gave equations that linked each other FWHM, peaks separation
and $v \sin i$ for objects whit EW of the H$\alpha$ emitting disk region
less than 15~{\AA}. From our measurements, we derived a velocity equal to 
$\approx$\,230~km~s$^{-1}$. 

The rotation law in Be star circumstellar disk is usually written as 
$v(r_d)~\propto~r_d^{-j}$, where the index {\it j}\,=\,0.5 if for keplerian 
rotation and {\it j}\,=\,1 for conservation of angular momentum.
According to former author, the observed normalized peaks separation is a 
function of the outer radius of the disk, expressed in term of
stellar radius, via the formula:

\begin{equation}
\frac{\Delta V}{2\,v \sin i}\,=\,r_d^{-j}
\label{vel}
\end{equation}

In their paper, \citet{hanu88} concluded that for their sample of Be stars the
value of {\it j} is closer to 1 than to 0.5. On the contrary, \citet{hummel00} 
concluded that the formation of H$\alpha$ shell profiles in Be stars generally
requires that {\it j}\,$\approx$\,0.5, i.e. the keplerian rotation is a valid 
approximation. As for our object, according to the Okazaki's model, the correlation
between RV and V/R showed in Fig.~\ref{vr_rv} is a strong argument for the validity 
of keplerian rotation. Thus, using Eq.~\ref{vel}, we estimated the radius of the 
outer disk surrounding $\beta$~Cep~Aa to be 3.38~R$_*$ ({\it j}\,=\,0.5).

\section{Conclusion}
In this paper we analyzed the last 13 years of the H$\alpha$ emission history of
$\beta$~Cep~Aa. Spectroscopic data, for a total of 72 spectra, have been
obtained both from {\it INAF - Osservatorio Astrofisico di Catania}
and from archives of other three observatories as described in Sec.~\ref{observ}.

We used H$\beta$ and H$\gamma$ profiles to derive effective temperature and
surface gravity of the primary. By spectral synthesis we obtained: 
T$_{\rm eff}$\,=\,24000\,$\pm$\,250 K and $\log g$\,=\,3.91\,$\pm$\,0.10.

Regarding the secondary, we conclude that $\beta$~Cep~Aa is a Be star which shows 
a high degree of variability in H$\alpha$ emission intensity as well as in the 
other line width parameters here considered. From 1993 to 1999 we observed a decrease 
in the emission and a subsequent increase till July 2002, that remains constant 
till August 2003. After this long cycle extended for 10 years, we again observed a 
decline toward a minimum and a raise toward a new phase of strong emission, in a 
shorter period of about 3 years.

As concerns, V/R ratio, peaks separation and radial velocity of central
depression, we also observed a long 10 years cycle out-of-phase with respect the EW
and a more rapid variability occured during the last 3 years. As for the FWHM, we
did not observe any appreciable change, at least at the resolving power of our data.

Making use of some useful literature correlations between FWHM, peaks separation 
and stellar projected rotational velocities, we estimated for the first time 
the $v \sin i$ of $\beta$~Cep~Aa equal to $\approx$\,230~km~s$^{-1}$
and the outer radius of the surrounding disk, $r_d$\,=\,3.38 R$_*$.

\section*{Acknowledgments}
This research has made use of the SIMBAD database, operated at CDS, Strasbourg, France. 
The author wish to thanks the anonymous referee for his/her 
suggestions useful to improve the scientific impact of the manuscript.

A warm thanks to Anna for her contribution in improving the english form 
of my original manuscript.

\appendix

\section{Journal of observations}

\begin{table}
\caption[]{Journal of observations. For each spectrum, ordered by heliocentric
julian date (HJD) we report: the equivalent width expressed in {\AA}, the ratio
between V and R peaks,
the peaks separations and the radial velocity of the central depression. The 
last two quantities are all expressed in km s$^{-1}$. The capital letter 
next to the HJD is a flag indicating the observatory: R - Ritter Observatory, 
I - INT, E - Elodie and O - Catania Astrophysical Observatory.}
\label{journal}
\begin{tabular} {lrccr}
\hline
\hline
~~~~~HJD    &     EW~~~~        & V/R             &  ~~~$\Delta$\,V &  RV$_{\rm CD}$            \\
(2400000.+) &  ({\AA})~~~~      &                 & (km s$^{-1}$)  & (km s$^{-1}$)               \\
\hline
49254.695 R & $-$11.5 $\pm$ 1.2 & 1.06 $\pm$ 0.10 & 256 $\pm$ 13 & 12.1 $\pm$ 1.4 \\
49261.590 R & $-$10.3 $\pm$ 1.3 & 1.07 $\pm$ 0.10 & 250 $\pm$ 10 & 18.6 $\pm$ 1.4 \\
49263.625 R & $-$11.3 $\pm$ 1.1 & 1.04 $\pm$ 0.08 & 255 $\pm$ 10 & 16.8 $\pm$ 1.1 \\
49266.582 R &  $-$9.9 $\pm$ 3.1 & 1.08 $\pm$ 0.10 & 248 $\pm$ 13 & 36.4 $\pm$ 2.0 \\
49268.583 R & $-$10.3 $\pm$ 4.5 & 1.04 $\pm$ 0.10 & 253 $\pm$ 12 & 14.2 $\pm$ 2.0 \\
49271.583 R & $-$10.9 $\pm$ 3.1 & 1.00 $\pm$ 0.10 & 250 $\pm$ 13 & 33.5 $\pm$ 2.3 \\
49319.551 R & $-$10.1 $\pm$ 4.1 & 1.06 $\pm$ 0.14 & 239 $\pm$ 16 & 37.0 $\pm$ 3.1 \\
49333.625 R &  $-$9.7 $\pm$ 0.8 & 1.04 $\pm$ 0.06 & 247 $\pm$  6 & 17.2 $\pm$ 0.5 \\
49334.658 R &  $-$9.4 $\pm$ 1.4 & 1.03 $\pm$ 0.08 & 248 $\pm$  7 & 20.8 $\pm$ 0.8 \\
49579.878 R &  $-$5.9 $\pm$ 1.7 & 1.09 $\pm$ 0.17 & 249 $\pm$ 12 & 24.1 $\pm$ 2.5 \\
49582.721 R &  $-$6.0 $\pm$ 2.4 & 1.05 $\pm$ 0.13 & 242 $\pm$ 11 & 23.0 $\pm$ 2.2 \\
49628.705 R &  $-$6.7 $\pm$ 1.5 & 1.09 $\pm$ 0.11 & 247 $\pm$  8 & 26.8 $\pm$ 1.1 \\
50030.278 E &  $-$4.8 $\pm$ 1.7 & 1.13 $\pm$ 0.14 & 250 $\pm$  8 & 39.8 $\pm$ 1.2 \\
50710.563 E &  $-$7.0 $\pm$ 0.9 & 1.14 $\pm$ 0.14 & 261 $\pm$ 17 & 51.7 $\pm$ 2.1 \\
50768.467 I &  $-$5.9 $\pm$ 0.4 & 1.20 $\pm$ 0.15 & 264 $\pm$ 11 & 42.3 $\pm$ 1.3 \\
50770.489 I &  $-$5.9 $\pm$ 2.8 & 1.16 $\pm$ 0.19 & 268 $\pm$ 15 & 52.4 $\pm$ 3.3 \\
51144.261 E &  $-$0.6 $\pm$ 0.6 & 1.79 $\pm$ 0.75 & 262 $\pm$ 26 & 58.8 $\pm$ 4.0 \\
51155.338 I &  $-$0.5 $\pm$ 0.7 & 2.27 $\pm$ 1.04 & 268 $\pm$ 32 & 76.9 $\pm$ 4.0 \\
51325.682 I &  $-$0.6 $\pm$ 1.2 & 1.66 $\pm$ 1.38 & 281 $\pm$ 28 & 72.6 $\pm$ 4.3 \\
52092.823 R & $-$11.3 $\pm$ 0.8 & 1.01 $\pm$ 0.09 & 260 $\pm$ 14 &  2.3 $\pm$ 1.7 \\
52104.862 R & $-$10.6 $\pm$ 0.6 & 1.02 $\pm$ 0.11 & 261 $\pm$ 14 &  0.9 $\pm$ 3.2 \\
52155.713 R &  $-$9.9 $\pm$ 0.5 & 1.03 $\pm$ 0.09 & 255 $\pm$ 10 &  2.8 $\pm$ 1.6 \\
52157.795 R &  $-$9.6 $\pm$ 0.4 & 1.05 $\pm$ 0.14 & 257 $\pm$ 16 &  0.6 $\pm$ 3.6 \\
52158.715 R &  $-$9.9 $\pm$ 1.1 & 1.04 $\pm$ 0.10 & 256 $\pm$ 12 & 11.0 $\pm$ 1.9 \\
52159.624 R &  $-$9.4 $\pm$ 3.3 & 1.02 $\pm$ 0.11 & 258 $\pm$ 12 & 22.3 $\pm$ 2.0 \\
52167.710 R &  $-$9.7 $\pm$ 0.5 & 1.05 $\pm$ 0.11 & 257 $\pm$ 11 &$-$2.4 $\pm$ 1.9\\
52220.580 R &  $-$9.6 $\pm$ 2.4 & 0.99 $\pm$ 0.11 & 259 $\pm$ 11 & 16.9 $\pm$ 1.9 \\
52223.565 R &  $-$9.6 $\pm$ 1.6 & 0.97 $\pm$ 0.12 & 241 $\pm$ 15 &  3.4 $\pm$ 3.0 \\
52861.537 E &  $-$9.6 $\pm$ 1.2 & 1.03 $\pm$ 0.12 & 255 $\pm$ 15 & 27.1 $\pm$ 2.0 \\
52862.436 E &  $-$8.6 $\pm$ 0.1 & 0.97 $\pm$ 0.11 & 262 $\pm$ 14 &$-$6.3 $\pm$ 2.0\\
52864.634 E & $-$10.2 $\pm$ 1.2 & 0.96 $\pm$ 0.10 & 258 $\pm$ 14 & 15.9 $\pm$ 1.8 \\
52871.460 E &  $-$9.3 $\pm$ 0.8 & 0.95 $\pm$ 0.10 & 251 $\pm$ 17 & 11.7 $\pm$ 2.0 \\
53240.765 R &  $-$2.2 $\pm$ 4.2 & 1.06 $\pm$ 0.36 & 255 $\pm$ 19 & 39.4 $\pm$ 5.4 \\
53250.731 R &  $-$1.3 $\pm$ 5.1 & 1.03 $\pm$ 0.26 & 240 $\pm$ 18 & 47.0 $\pm$ 5.4 \\
53261.625 R &  $-$2.0 $\pm$ 2.4 & 1.07 $\pm$ 0.33 & 255 $\pm$ 16 & 30.6 $\pm$ 3.8 \\
53264.599 R &  $-$0.8 $\pm$ 5.0 & 1.19 $\pm$ 0.49 & 261 $\pm$ 19 & 53.6 $\pm$ 5.3 \\
53271.686 R &  $-$2.6 $\pm$ 3.7 & 1.25 $\pm$ 0.36 & 251 $\pm$ 15 &  46.4 $\pm$  3.2\\ 
53276.663 R &  $-$2.2 $\pm$ 0.8 & 1.10 $\pm$ 0.24 & 256 $\pm$ 14 &  37.3 $\pm$  2.4\\ 
53278.598 R &  $-$2.8 $\pm$ 0.9 & 1.31 $\pm$ 0.34 & 243 $\pm$ 14 &  44.9 $\pm$  1.8\\ 
53279.627 R &  $-$2.5 $\pm$ 2.9 & 1.26 $\pm$ 0.34 & 238 $\pm$ 16 &  45.5 $\pm$  3.2\\ 
53281.697 R &  $-$2.9 $\pm$ 2.3 & 1.08 $\pm$ 0.18 & 230 $\pm$ 10 &  30.8 $\pm$  2.1\\ 
53282.588 R &  $-$2.1 $\pm$ 1.4 & 1.17 $\pm$ 0.42 & 250 $\pm$ 18 &  25.5 $\pm$  5.4\\ 
53284.703 R &  $-$2.9 $\pm$ 1.1 & 1.03 $\pm$ 0.44 & 240 $\pm$ 28 &  31.0 $\pm$  12.\\ 
53286.550 R &  $-$2.5 $\pm$ 0.6 & 1.10 $\pm$ 0.35 & 244 $\pm$ 17 &  36.1 $\pm$  3.5\\ 
53289.542 R &  $-$2.5 $\pm$ 0.5 & 1.08 $\pm$ 0.31 & 240 $\pm$ 15 &  29.5 $\pm$  2.7\\ 
53310.608 R &  $-$1.5 $\pm$ 0.9 & 1.14 $\pm$ 0.35 & 244 $\pm$ 15 &  32.9 $\pm$  2.4\\ 
53315.542 R &  $-$2.3 $\pm$ 0.3 & 1.12 $\pm$ 0.32 & 235 $\pm$ 15 &  34.5 $\pm$  2.4\\ 
53316.540 R &  $-$1.5 $\pm$ 1.0 & 1.17 $\pm$ 0.40 & 246 $\pm$ 17 &  36.5 $\pm$  2.6\\ 
53318.598 R &  $-$2.2 $\pm$ 2.2 & 1.09 $\pm$ 0.34 & 247 $\pm$ 15 &  24.1 $\pm$  2.7\\ 
53322.619 R &  $-$2.3 $\pm$ 2.3 & 1.25 $\pm$ 0.48 & 243 $\pm$ 19 &  25.4 $\pm$  3.3\\ 
53323.620 R &  $-$1.5 $\pm$ 4.5 & 0.92 $\pm$ 0.37 & 245 $\pm$ 21 &$-$12.2 $\pm$ 6.1\\ 
53324.524 R &  $-$2.5 $\pm$ 1.5 & 1.03 $\pm$ 0.38 & 252 $\pm$ 20 &  13.0 $\pm$  5.0\\ 
53336.500 R &  $-$1.4 $\pm$ 0.6 & 1.17 $\pm$ 0.40 & 244 $\pm$ 16 &  37.4 $\pm$  2.6\\ 
\hline     
\end{tabular}
\end{table}
\begin{table}
\contcaption{}
\begin{tabular} {lrccr}
\hline
\hline
~~~~~HJD    &     EW~~~~        & V/R             &  ~~~$\Delta$\,V &  RV$_{\rm CD}$            \\
(2400000.+) &  ({\AA})~~~~      &                 & (km s$^{-1}$)  & (km s$^{-1}$)               \\
\hline
53579.505 O &  $-$0.3 $\pm$ 1.4 & 0.87 $\pm$ 0.76 & 227 $\pm$ 35 &$-$29.3 $\pm$ 11.\\ 
53580.546 O &  $-$0.2 $\pm$ 0.7 & 0.87 $\pm$ 0.81 & 245 $\pm$ 30 &$-$38.1 $\pm$ 7.0\\ 
53581.559 O &     2.2 $\pm$ 1.9 & 0.78 $\pm$ 0.53 & 248 $\pm$ 29 &$-$35.4 $\pm$ 4.2\\ 
53635.525 O &     2.2 $\pm$ 1.2 & 0.98 $\pm$ 0.84 & 261 $\pm$ 34 &                 \\ 
53672.415 O &     1.7 $\pm$ 6.3 & 0.71 $\pm$ 0.61 & 264 $\pm$ 48 &                 \\ 
53673.457 O &  $-$1.5 $\pm$ 0.8 & 0.59 $\pm$ 0.31 & 266 $\pm$ 36 &                 \\ 
53674.475 O &     0.8 $\pm$ 4.7 & 0.84 $\pm$ 0.54 & 236 $\pm$ 51 &                 \\ 
53871.463 O &  $-$3.5 $\pm$ 0.7 & 0.96 $\pm$ 0.26 & 287 $\pm$ 20 &$-$18.3 $\pm$ 3.9\\ 
53872.511 O &  $-$4.6 $\pm$ 2.6 & 0.95 $\pm$ 0.35 & 298 $\pm$ 24 &$-$14.1 $\pm$ 4.6\\ 
53873.522 O &  $-$3.8 $\pm$ 2.0 & 1.03 $\pm$ 0.41 & 288 $\pm$ 26 & $-$5.9 $\pm$ 5.5\\ 
53933.850 O &  $-$3.6 $\pm$ 5.0 & 0.92 $\pm$ 0.27 & 273 $\pm$ 22 &$-$40.3 $\pm$ 3.3\\ 
54013.295 O &  $-$4.6 $\pm$ 5.0 & 0.97 $\pm$ 0.23 & 280 $\pm$ 21 &$-$19.8 $\pm$ 5.8\\ 
54016.544 O &  $-$5.3 $\pm$ 5.9 & 0.93 $\pm$ 0.20 & 296 $\pm$ 24 &$-$19.1 $\pm$ 5.2\\ 
54017.468 O &  $-$4.0 $\pm$ 6.1 & 0.94 $\pm$ 0.21 & 285 $\pm$ 22 &$-$18.4 $\pm$ 4.9\\ 
54018.399 O &  $-$5.5 $\pm$ 4.8 & 0.86 $\pm$ 0.18 & 289 $\pm$ 21 &$-$20.5 $\pm$ 4.7\\ 
54019.552 O &  $-$8.0 $\pm$ 4.5 & 0.89 $\pm$ 0.19 & 297 $\pm$ 18 &$-$23.1 $\pm$ 3.5\\ 
54021.544 O &  $-$5.6 $\pm$ 2.2 & 0.93 $\pm$ 0.17 & 275 $\pm$ 17 &$-$30.4 $\pm$ 2.9\\ 
54022.398 O &  $-$5.2 $\pm$ 5.5 & 1.08 $\pm$ 0.22 & 269 $\pm$ 17 &  16.9 $\pm$  3.5\\ 
54041.432 O &  $-$4.6 $\pm$ 3.0 & 0.96 $\pm$ 0.17 & 274 $\pm$ 16 &   6.8 $\pm$  2.9\\ 
\hline     
\end{tabular}
\end{table}

\label{lastpage}
\end{document}